# Coupling slot-waveguide cavities for large-scale quantum optical devices

Chun-Hsu Su,[1,2*] Mark P. Hiscocks,[3] Brant C. Gibson,[1] Andrew D. Greentree,[1]
Lloyd C. L. Hollenberg,[1,2] and François Ladouceur[3]

[1] *School of Physics, University of Melbourne, VIC 3010, Australia*
[2] *Centre for Quantum Computer Technology, School of Physics, University of Melbourne, VIC 3010, Australia*
[3] *School of Electrical Engineering and Telecommunications, University of New South Wales, NSW 2052, Australia*
[*]*chsu@ph.unimelb.edu.au*

By offering effective modal volumes significantly less than a cubic wavelength, slot-waveguide cavities offer a new in-road into strong atom-photon coupling in the visible regime. Here we explore two-dimensional arrays of coupled slot cavities which underpin designs for novel quantum emulators and polaritonic quantum phase transition devices.  Specifically, we investigate the lateral coupling characteristics of diamond-air and GaP-air slot waveguides using numerically-assisted coupled-mode theory, and the longitudinal coupling properties via distributed Bragg reflectors using mode-propagation simulations. We find that slot-waveguide cavities in the Fabry-Perot arrangement can be coupled and effectively treated with a tight-binding description, and are a suitable platform for realizing Jaynes-Cummings-Hubbard physics.

## 1. Introduction

Quantum emulation with controllable quantum systems [1] is an important field as it offers efficient means to study computationally-hard problems. The numerical roadblock arises because of the exponential growth in quantum systems with linear increase in the number of quantum particles, and so quantum emulation is expected to be important in many disciplines including atomic physics, quantum chemistry, condensed-matter physics, material engineering, high-energy physics and cosmology. There are now many platforms being explored for implementing quantum emulators using various quantum systems such as neutral atoms, ions, photons and electrons. These efforts are matched by fast and striking experimental advances and now the level of coherent controlling these systems required for the physical realization of quantum simulation is within reach [2].

Various models in condensed-matter physics to study highly-correlated many-body dynamics can be mimicked with extended cavity quantum-electrodynamics (QED) systems, namely two-dimensional (2D) networks of atoms in optical cavities in a Jaynes-Cummings-Hubbard model (JCH) with tight-binding inter-site interaction [3–5]. This approach of using artificial tunable systems offers much flexibility and control, owing to several controllable parameters such as nearest-neighbour tunneling, on-site interactions, multipartite interactions, multilevel atoms, feedback controls, and array geometry. Moreover, the ease of accessing each cavity individually in this implementation allows *in situ* measurements of local properties and correlation functions. By adjusting the on-site interaction energy and tunneling rate, nonlinear photon-photon repulsion and polaritonic quantum phase transition from a superfluid to a Mott-insulator phase can be generated and studied on a mesoscopic scale at ambience conditions [4]. However, in the optical regime, the crucial next step of building JCH-based quantum emulators is creating a 2D network of resonantly-coupled optical cavities that also supports strong intracavity atom-photon coupling. Notably, such systems would also open up new possibilities in the physics of the quantum Hall effect [6] and quantum metamaterials [7].

Photonic bandgap (PBG) cavities offer a promising route to large-scale solid-state cavity-QED applications. The reason for this derives from the great technological advancements in photonic structure fabrication and the ability to create wavelength-sized, high-$Q$ cavities [8]. In the appropriate media, these can be doped with impurities on the surface of the cavity, *e.g.* optical defect centres in diamond [9,10], donor-bound electrons [11] and quantum dots [12] in gallium-arsenide semiconductors. Prospects for large-scale integration in photonic crystals are promising, and over 100 PBG cavities have been coupled resonantly via a common line-defect to demonstrate ultraslow waveguiding [13].

Slot-waveguide cavities (SWCs) are another promising implementation [14,15]. Slot waveguides allow in principle lossless transmission along a narrow region (*i.e.* slot) defined by two regions (rods) of higher refractive index via large dielectric discontinuities. SWCs are then formed as a micro-ring or in a Fabry-Perot (FP) arrangement by combining a slot waveguide with mirrors, PBG or distributed Bragg reflectors (DBRs) [16]. A critical advantage of SWCs over PBG cavities in the quantum-emulator engineering context is that the cavity mode volumes of SWCs can be up to 2 orders of magnitude smaller [17–19], allowing stronger intracavity interactions, and thereby reducing the $Q$ limitations to demonstrate strong atom-photon coupling. As the maximum of the optical field is in the central low dielectric (ideally air) region, they are compatible with high-dipole moment emissive nanoparticles that can infiltrate the slot. In particular, we have explored coupling of the SWC to an optically-active defect centre in diamond [19] – negatively-charged nitrogen-vacancy (NV) centre – which is a promising quantum system for single-photon emission [20], quantum computing [21,22], quantum control [23,24], and communication [25].

To date, most investigations into SWCs have focused on their confinement properties for a single cavity. Here we consider the optical coupling between spatially separated identical SWCs in array configurations. We find that SWCs in the FP arrangement can be coupled laterally and end-to-end, and that in certain regimes this coupling can be effectively treated with a tight-binding type interaction. Through these characterizations, we also determine the DBR specifications for achieving subwavelength confinement in FP-based SWCs. Our results demonstrate the potential of SWCs for building large-scale integrated quantum-optical devices that take advantage of the ultra-small mode volumes of SWCs, and the availability of strong and coherent dipoles such as diamond optical centres, quantum dots and nanorods.

## 2. Slot-waveguide cavity array

In the tight-binding model, photons are tightly-confined within individual resonators and can propagate only by hopping between adjacent cavities via evanescent fields of one cavity tunneling into the other [26]. Here we want to achieve a device in which the interaction between cavity sites is effectively mimicking a tight-binding model and hence operating in the most commonly-discussed regime of quantum emulation.

For slot waveguides, the high refractive index contrast at rod-slot interfaces enables ideally lossless light guiding along the central slot. To be compatible with high-dipole moment emitters as part of the construction of cavity-QED systems in the visible regime, diamond and gallium phosphide (GaP) are suitable materials for the rods because they have high transparency at this wavelength range [27] and have high refractive indices ($n$ = 2.4 for diamond and 3.3 for GaP). On these structures, an air-slot and waveguiding patterns can be fabricated using electron beam [15] or optical lithography [28] combined with masking/dry etching techniques [29].

Light confinement along the $z$-direction (slot direction) to form a slot-waveguide cavity (SWC) is achieved by appending the ends of the waveguide with reflective boundaries such as mirrors, PBG or DBRs. Using narrow slots of $w_S$ = 20 nm width and an optimal $w_R \times h$ = 140 × 110 nm diamond rods, the fundamental quasi-TE modes of the structure occupies a cavity mode volume of $0.1\lambda^3/n$, where $\lambda$ is the operating wavelength of 637 nm and the mode is assumed to span over a cavity length of $\lambda/2$. The mode volume reduces to $0.02\lambda^3/n$ when replaced with 5 nm slot or a 110 × 70 nm GaP rods. Further reduction is possible using angled sidewalls for the slot [30,31]. Considering the special case of a NV in diamond with dipole moment on the zero phonon transition line of $10^{-30}$ Cm and emission wavelength 637 nm, the single-photon Rabi frequency $\Omega$, can in theory reach as high as $\Omega = 10^{11}$ rad/s [19]. This is an order of magnitude stronger than the intracavity, atom-photon coupling that can be realized by wavelength-sized PBG counterparts.

The critical next stage in integrated quantum device design is to move from isolated atom-cavity systems, to integrated quantum arrays. 1D and 2D arrays of coupled SWCs can be realized by placing them in parallel and in close proximity or by coupling adjacent SWCs using a common reflective boundary, as illustrated in Fig. 1. We consider two configurations for lateral inter-cavity coupling of the SWCs in Sec. 3. The first is the *cladding-separated* arrangement shown in Figs. 1(a,b) where the cavities are separated by a region of wide cladding, $w_G$. The other possible configuration is the *shared-rod* configuration depicted in Figs. 1(b,d), where there is no cladding. In each configuration, we will show regimes (where possible) that a tight-binding treatment can be used to model the interactions, and to thereby realize coupled cavity arrays. In Sec. 4, we investigate the end-to-end coupling between two slot cavities through partial transmission taking place at a shared DBR [Fig. 1(e)]. We investigate the lateral and longitudinal couplings with two different treatments. Specifically, the treatment of lateral coupling involves the numerical calculation of supermodes of slot-waveguide arrays in the assumption of weakly-coupled and infinitely long waveguides. Alternatively the longitudinal coupling via DBR is studied with mode-propagation simulations in a single SWC. In this analysis, we also explicitly consider the cavity mode extension into the DBRs and its effects on increasing the cavity mode volumes, which is a departure from the hard-boundary assumptions made in Ref. [19].

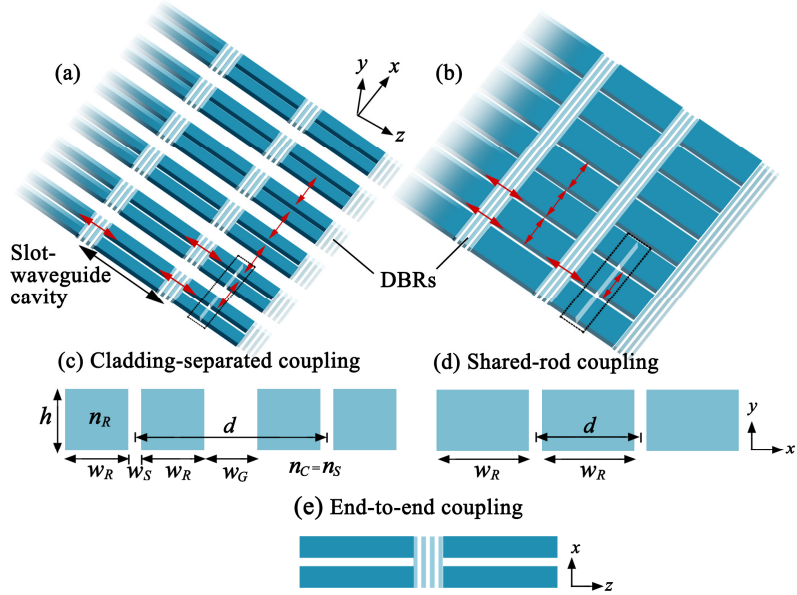

Fig. 1. Schematics of 2D coupled slot-waveguide cavity (SWC) arrays, where the lateral light transfer along *x* direction is realized through (a) a cladding-separated arrangement or (b) shared-rod arrangement. Their cross-sections in the *xy*-plane are depicted (c,d). (e) In both cases, the longitudinal coupling along *z* direction is realized by partial transmission at the distributed Bragg reflecting (DBR) boundary. Red arrows indicate nearest-neighbour couplings between cavities.

## 3. Lateral inter-cavity coupling

To begin our analysis of networks of coupled-cavity systems, we first consider a series of *N* identical, parallel slot waveguides separated by a distance *d*, aligned perpendicular to the long axis of the cavities (*x*-axis in Fig. 1). Coupled-mode theory (CMT) [32,33] allows the determination of the lateral transfer of light along the *x*-direction if the waveguides are weakly interacting. By working in the regime where the waveguides are sufficiently far apart and strongly guided in the tight-binding regime, the normal modes or the supermodes of the *N*-coupled array can then be approximated by an expansion of the magnetic $H_m$ and electric $E_m$ modal fields of the individual waveguides modes in isolation [34,35]. The coupling coefficient can be written down in terms of overlap integrals of the individual waveguide modes and refractive index profile of the arrayed structures in the *xy*-plane. In particular, when each waveguide supports only one TE-like mode in isolation, the total electric field distribution of a supermode is,

$$\mathbf{E}(x,y,z) = \sum_{m=1}^{N} A_m E_m(x,y) \exp(i\chi z) \hat{\mathbf{x}}. \tag{1}$$

where $\chi$ is the complex propagation constant of this supermode and $A_m$ is the modal amplitude. Using Eq. 1 as a trial solution, the wave equation can be recast as an eigen-equation in matrix form [34]: **MA = BA**, where $\mathbf{A} = [A_1\ A_2\ \ldots\ A_N]^T$, and **B** is a diagonal matrix $\chi^2 \mathbf{I}$. The elements of the coupling matrix **M** are

$$M_{mn} = \begin{cases} \beta_m^2 + \kappa_{mn} & , m = n \\ (\beta_n^2 - \chi^2)I_{mn} + \kappa_{mn} & , m \neq n \end{cases}, \quad (2)$$

where $\beta_m$ ($m = 1, 2, \ldots N$) is the propagation constant of the $n$th waveguide in isolation and $\beta_m = \beta$ for identical guides, $\kappa_{mn}$ is the mutual coupling coefficient between $m$th and $n$th guides ($\kappa_{nn}$ is the self-coupling coefficient), and $I_{mn}$ denotes the overlap of the non-orthogonal modes of any two guides. The mutual coupling is determined by the integrals over the entire cross section,

$$\kappa_{mn} = k^2 \sum_{l=1, l \neq n}^{N} \iint [n_l(x,y)^2 - n_{cl}^2] E_m^*(x,y) E_n(x,y) dxdy \quad (3)$$

with $n_l$ being the refractive index of $l$th waveguide in absence of the others, and $n_{cl}$ the index of the cladding. Hence **M** provides insights into the orthogonality of the modes and the relative strengths of the inter-guide couplings. Cooper and Mookherjea [34] have proposed a method known as numerically-assisted CMT (NA-CMT) that uses exact numerical results of finite-difference frequency-domain mode solver to back-calculate **M** using the relation

$$\mathbf{M} = \mathbf{AXA}^{-1} \quad (4)$$

where $\mathbf{X} = \text{diag}\{\chi_m^2\}$ is a diagonal matrix in terms of the propagation constant $\chi_m$ of the $m$th supermodes. The purpose of this approach is to allow one to compare the exact scattering matrix with Eq. 2 to ascertain that the assumption of nearest-neighbour coupling is valid. In Sec. 3.1 and 3.2, we employ their approach to investigate TE supermodes of parallel slot-waveguide arrays. We used FIMMWAVE [36] to determine their modal amplitudes and propagation constants.

To describe the system in a tight-binding model, it is useful to determine the single-photon hopping or coupling rate $J$ (coupling energy $\hbar J$). Given that the solution of classical Maxwell's equations can be reinterpreted as a precise quantum description of a one-photon state, we use the classical results and write down the relation for lateral coupling,

$$J_L = \frac{c \kappa_{mn}}{2 n_{\text{eff}} k} \quad (5)$$

where the subscript $L$ is used to distinguish lateral coupling from end-to-end coupling ($J_E$) discussed in Sec. 4. The propagating field sees an effective refractive index $n_{\text{eff}}$ of the combined system that we approximate to that of an individual waveguide in isolation.

*3.1 Cladding-separated configuration*

We use the optimal diamond-air and GaP-air vertical slot arrangements previously discussed in Sec. 2 and first consider waveguide arrays with varying width of the separating cladding region. We first consider a two-waveguide system in Fig. 2, showing its even and odd supermodes (with respective propagation constants $\chi_+$ and $\chi_-$) and a strong E-field within each slot region. In the limit of large separations, the effective indices and propagation constants of these modes approach the parameter values of the single waveguide $n_{\text{eff}} = 1.31$ and $\beta = 12.91$ rad/µm for diamond-air, and 1.34 and 13.22 rad/µm for GaP-air slots. Applying NA-CMT analysis to the mode solutions of the diamond-air system with centre-to-centre separation distance $d = 500$ nm, we arrive at a matrix with $M_{nn} - \beta^2 = 0.1460$ and $M_{mn} = 2.6442$, in units of rad$^2$/µm$^2$. The self-coupling is much weaker than the inter-cavity coupling. In particular, with $\chi_+ > \beta > \chi_-$, $\kappa_{mn} = M_{mn}$ with $J_L = 3.1 \times 10^{13}$ rad/s for a centre-to-centre separation distance of $d = 500$ nm. This rate falls off exponentially with distance as shown in Fig. 3 and the contribution of $(M_{nn}-\beta^2)$ approaches zero. At 1 μm separation, $J_L = 2.4 \times 10^{10}$ rad/s is commensurate with the strength of the intracavity coupling of diamond colour centres to subwavelength-sized SWCs [19] On the other hand, for $d < 400$ nm, the cladding region between the waveguides begins to act as a slot, altering the number of effective cavities. The resulting geometry becomes akin the shared-rod configuration, which as we will see in Sec. 3.2 is not suitable for realizing tight-binding systems.

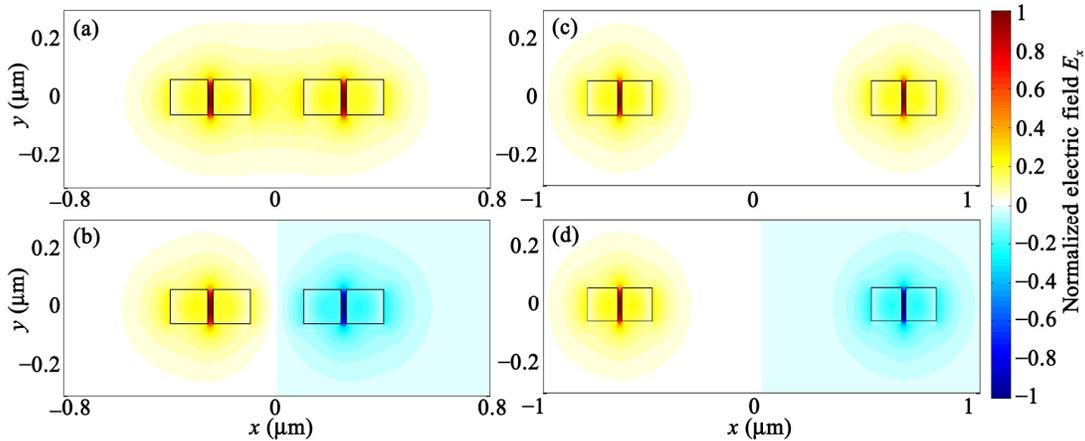

Fig. 2. $E_x$-field distribution of the TE-supermodes of a coupled diamond-air slot-waveguide system with dimensions $\{w_S, w_R, h\} = \{20, 140, 110\}$ nm for (a,b) $w_G = 200$ nm or $d = 500$ μm) and (c,d) 1 μm ($d = 1.3$ μm). Strong E-field localization is found within the slots. The even supermodes are shown in (a,c) and the odd in (b,d). The predicted coupling strength between the waveguides is shown in Fig. 3.

Next we apply NA-CMT to a larger array of 5 slot waveguides to study the non-nearest neighbour coupling between waveguides. Since such coupling is most prominent for small $d$, we focus on the diamond-air case of $d = 500$ nm. The cross-sectional mode distributions of its supermodes are shown in Fig. 4 and the associated coupling matrix is,

$$\mathbf{M} = \begin{pmatrix} 0.1574 & 2.6420 & -0.1775 & 0.0196 & -0.0025 \\ 2.6284 & 0.3032 & 2.6262 & -0.1756 & 0.0191 \\ -0.1754 & 2.6267 & 0.3046 & 2.6263 & -0.1757 \\ 0.0192 & -0.1761 & 2.6269 & 0.3041 & 2.6276 \\ 0.0026 & 0.0196 & -0.1780 & 2.6427 & 0.1560 \end{pmatrix} + \beta^2 \mathbf{I}, \quad (6)$$

with the physical meaning of each matrix element given in Eq. 2. The $M_{nn}$ and $M_{n,n\pm1}$ terms agree with the values of the nearest-neighbour coupling and self-coupling in the two-waveguide system. The near symmetry of the matrix suggests that the fields inside the individual waveguides remain centered within the slots (Fig. 4) such that it is still accurate to apply CMT and read off the peak amplitudes of the supermodes as the eigenvectors $\mathbf{A}$ [34]. Importantly, the nearest-neighbour coupling between the waveguides is the most dominant interactions by at least an order of magnitude. Further we find that the ratios of $\kappa_{nn}/\kappa_{n,n\pm1}$, $\kappa_{n,n+2}/\kappa_{n,n+1}$ decrease rapidly with the separation distance. As a comparison we find that at $d = 600$ nm matrix asymmetry reduces, the effective indices of the supermodes converge, $\kappa_{22}/\kappa_{n,n\pm1} \approx 2\kappa_{11}/\kappa_{n,n\pm1} \approx 0.04$, and $\kappa_{n,n+2}/\kappa_{n,n\pm1} = 0.03$. Thus, accurate implementations of 1D tight-binding models with dominant nearest-neighbour interactions and negligible non-nearest neighbour coupling can be achieved by setting up slot-waveguide arrays in this arrangement with modest guide separations.

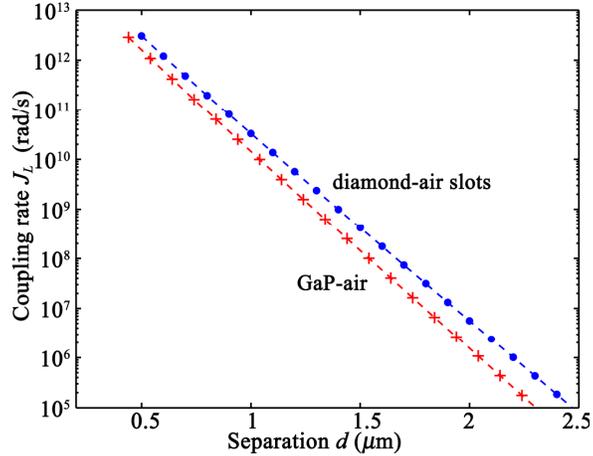

Fig. 3. Lateral coupling rate $J_L$ between two slot waveguides in cladding-separated configuration, plotted as a function of centre-to-centre separation between the waveguides. Circles (Crosses) denote the calculated results for coupled diamond-air (GaP-air) slots. The coupling strength exponentially decreases with the separation.

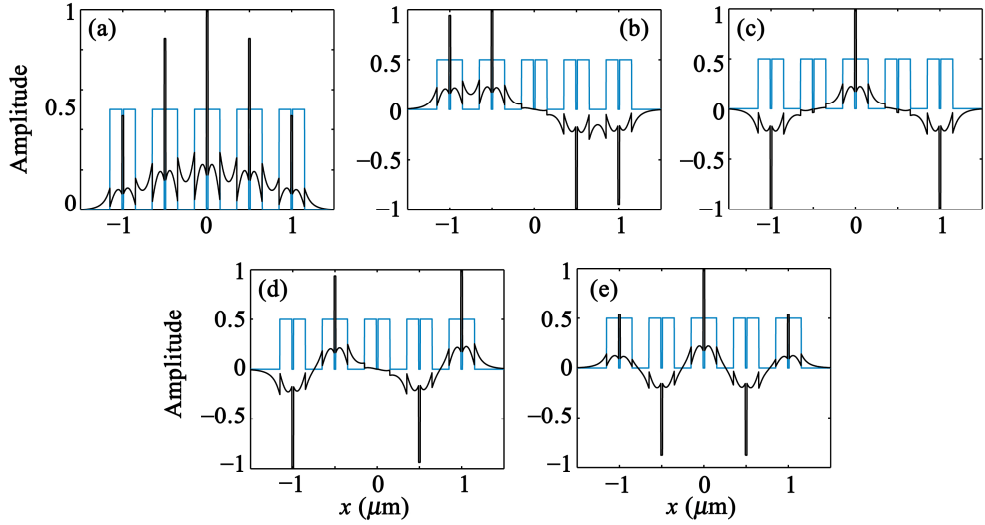

Fig. 4. Cross-sectional $E_x$-field distribution (black curves) of the TE-modes of a five coupled diamond-air slot-waveguides (blue) with separated by a 200 nm wide cladding (*i.e.* $d = 500$ nm). The M-matrix for this structure is given in Eq. 6. The SWC dimensions follow Fig. 2(a).

*3.2 Shared-rod configuration*

It is also useful to achieve stronger lateral coupling in a shared-rod arrangement [Fig. 1(b,d)] without an inter-site cladding region. However, in the two-waveguide setup, we find that there is a very limited range of waveguide separation where the slots remain guiding. In particular when the waveguides are too close together ($w_R < 100$ nm), the effective refractive index of the odd supermode falls below 1 and the mode is no longer guiding. As the width of the separating rod increases beyond 200 nm, it accommodates an increasingly larger E-field and the confinement capability of the slots diminishes. Following the prescription of Eqs. 4 and 5, we plot in Fig. 5(a) the inter-cavity coupling versus $d$, showing that such an arrangement only enables very strong coupling of order $10^{14}$ rad/s for a narrow range of rod thickness. We also find that the associated self-coupling term $M_{nn}$ increases almost linearly with separation from 130 rad$^2$/μm$^2$ by 2–3 fold in the plotted range, which is different from the typical behaviour in cladding-separated setup where the self-coupling decreases with separation. These observations can be explained as follows. In this analysis, we have assumed a fixed rod height $h$ while modifying the rod width. This has the effect of reducing the slot confinement property to produce larger modal overlap and allowing the central rod to accommodate a mode as we deviate from the optimal rod specifications that maximize intracavity field amplitude [19].

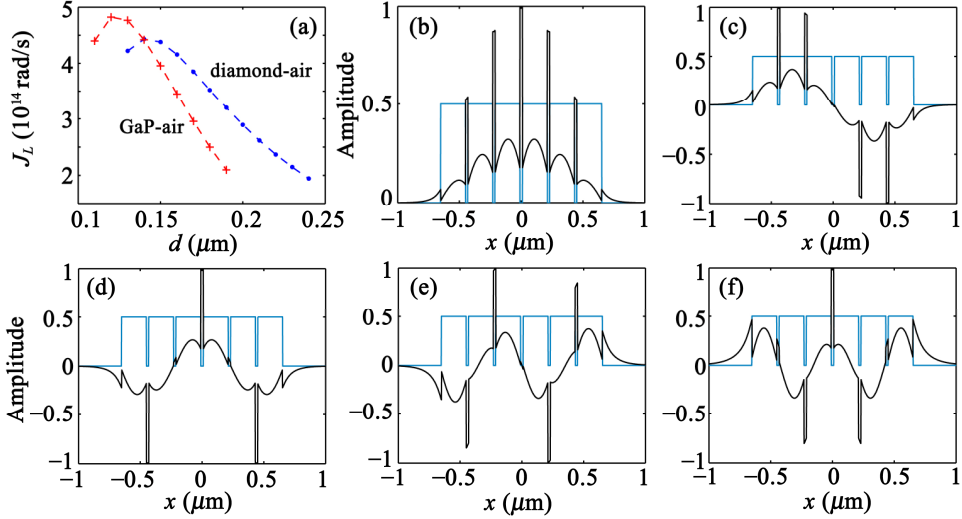

Fig. 5. (a) Lateral coupling rate $J_L$ between two different slot waveguides in shared-rod configuration. (b–f) E-field distribution of the five supermodes of a 5-waveguide array with separation $d$ of 220 nm. The M-matrix for this structure is given in Eq. 7.

We now turn to the modal calculation of a 5-waveguide array. We find that structures with small rod widths are unable to support 5 modes. Therefore we are limited to calculating the coupling matrices for the range $170 \leq w_R \leq 240$ nm and $140 \leq w_R \leq 240$ nm for diamond-air and GaP-air slots, respectively. As a representative example, we write down the coupling matrix for the case of $d = 220$ nm, with its supermodes shown in Fig. 5(b–f),

$$\mathbf{M} = \begin{pmatrix} 241.578 & 21.076 & 0.820 & -2.350 & 1.309 \\ 27.712 & 228.497 & 29.854 & -5.124 & 1.168 \\ -5.026 & 30.684 & 226.753 & 30.678 & -5.102 \\ 1.184 & -5.198 & 29.803 & 228.677 & 27.676 \\ 1.201 & -2.110 & 0.429 & 21.461 & 241.212 \end{pmatrix}. \quad (7)$$

In comparison with the cladding-separated array (Eq. 6), we see that non-nearest neighbour coupling and mode non-orthogonality are relatively strong and $\kappa_{nm}$ does not diminish with increasing separation. Given the fact that there is a pronounced matrix asymmetry, it is no longer accurate to use its modal peak amplitudes as the eigenvectors and the CMT approach. These undesirable features of coupling characteristics, field skewing inside the cavities, and limited operating range, suggest that this setup is not suitable for implementing practical tight-binding systems.

## 4. Longitudinal end-to-end coupling

The longitudinal coupling along the $z$-direction between two SWCs can be realized through reflection and transmission at a shared mirror. In this study, we consider the use of DBRs which may be added to the lithography step used to pattern the slot waveguides. From the reflectivity of the DBR and the average round-trip time we can calculate the average photon-hopping rate. Specifically we first note that the effective length contribution $L_\text{eff}$ of each grating to the total length $\hat{L}_c$ of the cavity mode in the $z$-direction is given by [37],

$$L_{\text{eff}} = L_{\text{gr}} \frac{\sqrt{R}}{2\tanh^{-1}(\sqrt{R})} \qquad (8)$$

where $L_{\text{gr}}$ is the grating physical length and $R$ is its reflection coefficient. The equation has the property that as the reflection tends to unity the length $L_{\text{eff}}$ tends to zero, and the maximum contribution of the grating is $L_{\text{gr}}/2$ at $R = 0$. Consequently, the adjusted modal length is $\hat{L}_c = L_c + 2L_{\text{eff}}$ where $L_c$ is the length of the SWC from the end of one DBR to the start of the next. As an extension to the cavity mode calculations in Ref. [19] that assumed hard or metallic boundaries with unit $R$. we revise the equation for cavity mode volumes to include field penetration into Bragg regions,

$$V = \frac{\int n(\mathbf{r})^2 |\mathbf{E}(\mathbf{r})|^2 \, d^2\mathbf{r}}{n(\mathbf{r}_{\max})^2 |\mathbf{E}(\mathbf{r}_{\max})|^2} \int_0^{\hat{L}_c} \sin^2(2\pi z/\hat{L}_c) dz \qquad (9)$$

where the length integral along the $z$-direction is now computed over the adjusted length $\hat{L}_c$. The area integral over the $xy$ cross section of the cavity uses the E-field distribution $\mathbf{E}(\mathbf{r})$ of the mode solution, and is scaled with respect to the position $\mathbf{r}_{\max}$ where the product $n(\mathbf{r})^2|\mathbf{E}(\mathbf{r})|^2$ is maximum. The end-to-end coupling rate $J_E$ is equivalent to the fractional loss of cavity field per round-trip time of $\tau = \hat{L}_c N_{\text{eff}}/c$, where $N_{\text{eff}}$ is the effective index of the combined systems of the slot waveguide and two DBRs. This coupling rate is related to the transmission $T$ and reflection $R$ coefficients via, for $T \ll 1$

$$J_E/2\pi \approx \frac{1-\sqrt{R}}{\tau} = \frac{T}{2\tau}. \qquad (10)$$

To estimate the grating length necessary for achieving high reflectivity, we use FIMMPROP [38] to simulate mode-propagation along a single SWC in its fundamental TE-mode to determine the reflected power at a DBR. As we require high reflection over short lengths, gratings that use small perturbations to the slot-waveguide dimensions, e.g., an increase in rod width $w_R$ of 10 nm as in Ref. [39], are insufficient, requiring more than 200 periods to achieve reflectivity more than 90%. Instead we use a high-contrast grating as in Ref. [40] for example, in our case leaving the material slab unetched to give a large index contrast between the air slot and rectangular rod, and use a 50/50 duty cycle, as shown in

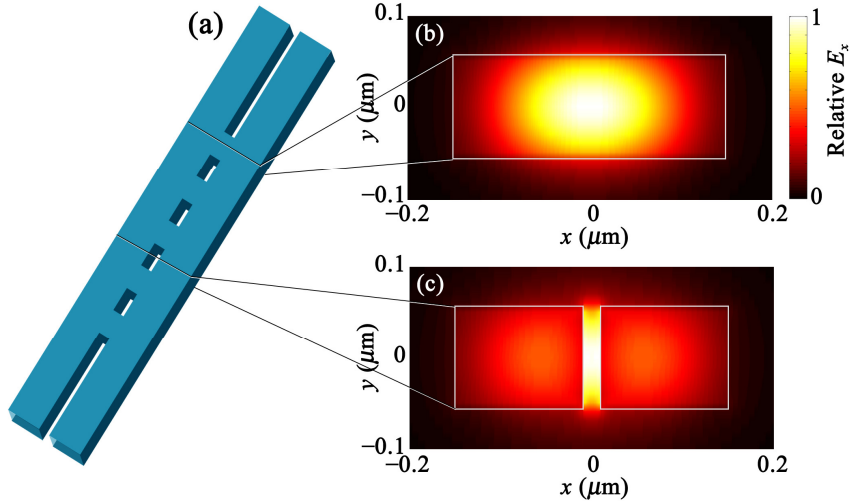

Fig. 6. (a) Schematic for a distributed Bragg reflector with 4.5 periods that separates two SWCs. The grating alternates between solid rectangular guide and slot regions. The respective E-field distributions of their fundamental modes are shown in (b) and (c).

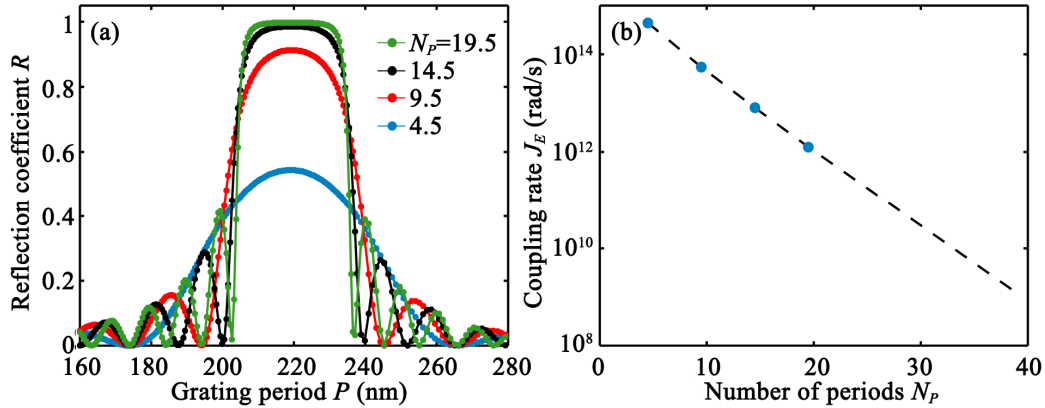

Fig. 7. (a) Reflection coefficient $R$ of a diamond-air DBR [Fig. 6(a)] appended to a diamond-air slot waveguide. The coefficient is associated to its fundamental TE-mode and is calculated for different grating periods $P$ and number of periods $N_p$. The maximal reflectivity is observed at $P = 220$ nm and these specifications are selected for (b) plotting longitudinal end-to-end coupling rate $J_E$. In both figures, markers indicate simulation data points, and the lines are guides and are used to predict the coupling rate at larger $N_p$.

Fig. 6. Due to the high contrast and nanometre-sized geometries involved, the simulations yield good indications of the reflection strengths, although the raw output shows numerical artifacts where reflected power slightly exceeds unity. To remove these artifacts, we invoke the conservation relation that $T + R = 1$ to plot the renormalized reflection spectra for different number of grating periods $N_P$ in Fig. 7(a). The maximum reflection occurs at the grating period $P$ of 220 nm that satisfies the relation that $P = \lambda/2n_{DBR}$ where we estimate $n_{DBR} = 1.45$ using the weighted average of the diamond-air slot (with an effective refractive index of 1.31) and diamond core (index 1.6).

Given these results, we expect that a grating with 14.5 periods and length 3.18 μm produces a high reflectivity of $1 - R = 10^{-2}$ and an effective grating length of $L_{eff} = 560$ nm. By increasing the number of periods to 19.5, the reflectivity improves to $1 - R = 10^{-3}$ while $L_{eff}$ hardly changes (570 nm). It is therefore worthwhile to point out that the effective length remains relatively constant around 0.6 μm because the denominator of Eq. 8 increases very slowly with increasing $R$. Consequently, since the minimum value of the cavity length $L_c$ is limited only by the size of the atomic system inside the cavity, we can set the overall index $N_{eff} = n_{DBR}$ and $\hat{L}_c \approx 2L_{eff}$ as the fundamental limit. Applying these values to Eqs. 8 and 10 we plot the expected coupling rate for different grating periods in Fig. 7(b). Apart from showing an exponential scaling, achieving a coupling strength in the range of $10^9$–$10^{11}$ rad/s is possible with structures with $24.5 < N_P < 39.5$ and $200 < P < 240$ nm. Moreover, following Eq. 9, we can expect that a dielectric-based grating design would only increase the cavity mode volume of a $\lambda/2$-long SWC reported in Ref. [19] by a factor of $(2L_{eff})/(\lambda/2) = 3.7$. The resultant cavity mode volumes are therefore still well below the dimensions of the wavelength of light.

## 5. Conclusions

The use of slot-waveguide cavities for implementing multi-cavity quantum-optical devices takes advantage of the extremely strong atom-photon interactions inside the cavities made possible by their subwavelength-sized cavity mode volumes. This effort to study and optimize the confinement properties of such cavities in various materials and designs is matched by our effort to investigate the coupling properties between multiple cavities arranged in parallel and in series. Amongst the three coupling configurations considered in this work, the arrangements of cladding-separated slots in parallel and DBR-separated slots connected end-to-end are ideal for implementing a tight-binding model.

We have shown that the strength of inter-cavity coupling in the lateral direction can be tailored from as large as $10^{12}$ rad/s to arbitrarily small values by introducing cladding separation of 500 nm or more.

Similarly, the coupling rate in the longitudinal direction can be reduced from $10^{14}$ rad/s by introducing 5 or more Bragg periods. Specifically, there is exponential scaling of coupling strength against these setup parameters. Our DBR analysis furthermore extends and validates the work of Ref. [19] by relaxing the hard-boundary assumption used in the calculation of cavity mode volumes of diamond-air and GaP-air slot-waveguide cavities.

These results, which along with Refs. [17,19], represent a blueprint for building a cavity-QED based quantum simulator to study strongly-correlated many-body systems. A very wide parameter space for the Hubbard-like inter-site interactions in JCH model is experimentally accessible in the proposed platform. For instance, the second-excitation Mott-insulator lobe of the quantum phase in JCH model can be physically simulated in a NV-doped cavity-array that supports a uniform inter-cavity coupling of $10^9$ rad/s [4]. This can be achieved with a lateral slot separation of 1.2 μm and 35 Bragg periods for the DBRs. More generally, our results support the suitability of slot designs as a promising and superior alternative to PBG cavities in the quantum-optical applications and large-scaled solid-state cavity-QED.

## Acknowledgments

This project is supported by the Australian Research Council under the Discovery Scheme (DP0770715, DP0880466, and DP0877871), and the Centre of Excellence Scheme (CE0348250). CHS acknowledges the support of the Albert Shimmins Memorial Fund. The authors thank Photon Design support for their assistance with simulations.